\documentclass[aps,prl,twocolumn,superscriptaddress,showpacs]{revtex4}

\usepackage{mathrsfs}
\usepackage{graphicx}
\usepackage{color}
\usepackage{amsmath}
\usepackage{amssymb}

\begin{document}
 
\title{Scaled Brownian motion: a paradoxical process with a time dependent
diffusivity for the description of anomalous diffusion}

\author{Jae-Hyung Jeon}
\affiliation{Department of Physics, Tampere University of Technology, FI-33101
Tampere, Finland}
\author{Aleksei V. Chechkin}
\affiliation{Institute for Theoretical Physics, Kharkov Institute of Physics
and Technology, Kharkov 61108, Ukraine}
\affiliation{Max-Planck Institute for the Physics of Complex Systems,
N{\"o}thnitzer Stra{\ss}e 38, 01187 Dresden, Germany}
\affiliation{Institute of Physics \& Astronomy, University of Potsdam, 14776
Potsdam-Golm, Germany}
\author{Ralf Metzler}
\email{rmetzler@uni-potsdam.de}
\affiliation{Institute of Physics \& Astronomy, University of Potsdam, 14776
Potsdam-Golm, Germany}
\affiliation{Department of Physics, Tampere University of Technology, FI-33101
Tampere, Finland}

\begin{abstract}
Anomalous diffusion is frequently described by scaled Brownian motion (SBM), a
Gaussian process with a power-law time dependent diffusion coefficient. Its mean
squared displacement is $\langle x^2(t)\rangle\simeq\mathscr{K}(t)t$ with
$\mathscr{K}(t)\simeq t^{\alpha-1}$ for $0<\alpha<2$. SBM may
provide a seemingly adequate description in the case of unbounded diffusion, for
which its probability density function coincides with that of fractional Brownian
motion. Here we show that free SBM is weakly non-ergodic but does not exhibit a
significant amplitude scatter of the time averaged mean squared displacement.
More severely, we demonstrate that under confinement, the dynamics encoded by SBM
is fundamentally different from both fractional Brownian motion and continuous
time random walks. SBM is highly non-stationary and cannot provide a physical
description for particles in a thermalised stationary system. Our findings have
direct impact on the modelling of single particle tracking experiments, in
particular, under confinement inside cellular compartments or when optical
tweezers tracking methods are used.
\end{abstract}


\maketitle

The passive, thermally driven diffusion of fluorescently labelled molecules or
optically visible submicron particles is routinely measured inside living cells
and complex liquids by methods such as single particle tracking \cite{spt} and
fluorescence correlation spectroscopy \cite{fcs}. Often the observed diffusion
patterns deviate from the laws of Brownian motion and display anomalous diffusion
characterised by the mean squared displacement (MSD)
\begin{equation}
\label{msd}
\langle x^2(t)\rangle\simeq K_{\alpha}t^{\alpha}
\end{equation}
with the anomalous diffusion coefficient $K_{\alpha}$ of dimension $\mathrm{cm}^2/
\mathrm{sec}^{\alpha}$. Depending on the magnitude of the anomalous diffusion
exponent one distinguishes subdiffusion ($0<\alpha<1$) and superdiffusion ($1<
\alpha<2$) \cite{report,hoefling,pt}.

In biological contexts one often invokes the concept of apparent subdiffusion,
which is in fact a transient crossover between free normal diffusion with $\alpha
=1$ and the thermal plateau of the MSD in confinement \cite{saxton}. However,
there is clear experimental evidence of long-ranged subdiffusion (\ref{msd}) in
living biological cells \cite{golding,weber,garini,weigel,tabei,lene,hoefling} and
in artificially crowded or structured liquids \cite{lene1,szymanski,pan,weiss_pccp,
skaug}. The physical
mechanisms for such anomalous diffusion are very varied, even considering the
most common approaches: we mention the continuous time random walk approach
\cite{scher,report,pt}, in which subdiffusion is effected by multiple trapping
or sticking events with an emerging power-law distribution of waiting times, as
seen in the data of Refs.~\cite{lene,tabei,weigel}. Alternatively, subdiffusion
arises in fractal geometries with their dead ends and bottlenecks across spatial
scales \cite{fractal}, an additional mechanism pinpointed in Ref.~\cite{weigel}.
We also mention diffusion processes in which the anomaly arises from the spatial
dependence of the diffusion coefficients \cite{andrey,andrey_pccp,new_prl}, or
from interaction of the tracer particle with a structured environment
\cite{aljaz_gel,max_pccp,wong,xu}.

A special role play the Gaussian models for anomalous diffusion which belong to
two families, that can best be distinguished on the Langevin equation level.
The Mandelbrot-van Ness fractional Brownian motion (FBM) follows the stochastic
equation $\dot{x}(t)=\zeta_{\mathrm{fGn}}(t)$,
where $\zeta_{\mathrm{fGn}}(t)$ represents fractional Gaussian noise characterised
by the covariance $\langle\zeta_{\mathrm{fGn}}(t_1)\zeta_{\mathrm{fGn}}(t_2)\rangle
\simeq\alpha(\alpha-1)|t_1-t_2|^{\alpha-2}$ which has long-ranged negative or
positive correlations, respectively,
for subdiffusion ($0<\alpha<1$) or superdiffusion ($1<
\alpha<2$) \cite{mandelbrot}. FBM and the closely related fractional Langevin
equation motion are physically related to the motion in a viscoelastic environment
\cite{goychuk}. In various experiments this type of motion was identified as single
or partial component \cite{lene,tabei,guigas,garini,lene1}.

Here we focus on the second family of Gaussian anomalous diffusion models, namely,
scaled Brownian motion (SBM) \cite{lim}. SBM is governed by the Langevin equation
\begin{equation}
\label{langevin}
\dot{x}(t)=\sqrt{2\mathscr{K}(t)}\times\zeta(t),
\end{equation}
where $\zeta(t)$ is white Gaussian noise with normalised covariance $\langle
\zeta(t_1)\zeta(t_2)\rangle=\delta(t_1-t_2)$. The prefactor in Eq.~(\ref{langevin})
is the power-law time dependent diffusion coefficient
\begin{equation}
\mathscr{K}(t)=\alpha K_{\alpha}t^{\alpha-1},
\end{equation}
which decreases or increases in time for $0<\alpha<1$ and $1<\alpha<2$,
respectively. SBM is used to model anomalous diffusion in a wide range of
systems \cite{weiss,verkman,wu,szymaski,mitra,lutsko}, in particular, in FRAP
experiments \cite{saxton1}. In terms of the time dependent diffusivity,
the MSD (\ref{msd}) can be rewritten as $\langle x^2(t)\rangle\simeq\mathscr{
K}(t)\times t$. Originally, the diffusion process with a time dependent 
diffusivity was introduced by Batchelor in the description of relative
diffusion in turbulence \cite{batchelor}.

For unconfined SBM the probability density function
is the Gaussian \cite{lim}
\begin{equation}
\label{prop}
P(x,t)=\sqrt{\frac{1}{4\pi K_{\alpha}t^{\alpha}}}\times\exp\left(-\frac{x^2}{4
K_{\alpha}t^{\alpha}}\right).
\end{equation}
It has exactly the same form as the free PDF of FBM \cite{lim}, if only
both processes are started at the origin. Despite
this deceiving similarity, we show that SBM is fundamentally different from
FBM and all other anomalous diffusion models mentioned above. Most importantly,
we show that it cannot represent a physical model for anomalous diffusion in
stationary or thermalised systems. Concurrently, the
remarkable properties of SBM revealed here may have other relevant applications,
as briefly discussed below.

\section{Time averaged mean squared displacement for unconfined SBM}

While we are used to think of a diffusion process in terms of the MSD (\ref{msd})
calculated as the spatial average of $x^2$ over the probability density function
$P(x,t)$, single particle tracking experiments typically provide few but long
individual time series $x(t)$. These are evaluated in terms of the time averaged
MSD \cite{pt}
\begin{equation}
\label{tamsd}
\overline{\delta^2(\Delta,t)}=\frac{1}{t-\Delta}\int_0^{t-\Delta}\Big(x(t'+\Delta)
-x(t')\Big)^2dt',
\end{equation}
where $t$ is the length of the time series (measurement time) and $\Delta$ the
lag time. In an ergodic system for sufficiently long $t$ ensemble and time averages
provide identical information, formally, $\langle x^2(\Delta)\rangle=\lim_{t\to
\infty}\overline{\delta^2(\Delta,t)}$, and $\overline{\delta^2}$ for different
trajectories are identical. For anomalous diffusion the behaviour of the MSD
(\ref{msd}) and the time averaged MSD (\ref{tamsd}) may be fundamentally different.
Simultaneously $\overline{\delta^2(\Delta,t)}$ for different trajectories may become
intrinsically irreproducible. This phenomenon of irreproducibility and disparity
$\langle x^2(\Delta)\rangle\neq\lim_{t\to\infty}\overline{\delta^2(\Delta,t)}$ is
usually called weak ergodicity breaking \cite{pt} and was discussed in detail for
the subdiffusive CTRW \cite{yonghe,lubelski,pccp,pnas,igor,pt}. In the following,
for simplicity we neglect the explicit dependence of $\overline{\delta^2}$ on the
measurement time $t$.

For SBM the mean of the time averaged MSD over multiple trajectories $i$, $\left<
\overline{\delta^2(\Delta)}\right>=N^{-1}\sum_{i=1}^N\overline{\delta^2_i(\Delta)}$
can be derived from the Langevin equation (\ref{langevin}), yielding
\cite{fulinski1}
\begin{equation}
\label{tamsd_sbm_full}
\left<\overline{\delta^2(\Delta)}\right>=\frac{2K_{\alpha}t^{1+\alpha}}{
(\alpha+1)}\frac{\left[1-\left(\frac{\Delta}{t}\right)^{1+\alpha}-\left(1-
\frac{\Delta}{t}\right)^{1+\alpha}\right]}{t-\Delta}.
\end{equation}
In the limit $\Delta\ll t$, we recover the behaviour \cite{thiel}
\begin{equation}
\label{tamsd_sbm}
\left<\overline{\delta^2(\Delta)}\right>\sim2K_{\alpha}\frac{\Delta}{t^{
1-\alpha}},
\end{equation}
and when the lag time $\Delta$ approaches the measurement time $t$ the limiting
form
\begin{equation}
\left<\overline{\delta^2(\Delta)}\right>\sim2K_{\alpha}t^{\alpha}-
\frac{\alpha K_{\alpha}}{t^{1-\alpha}}(t-\Delta)+\frac{\alpha(\alpha-1)
K_{\alpha}}{3t^{2-\alpha}}(t-\Delta)^2
\end{equation}
describes a cusp around $\Delta=t$. Result (\ref{tamsd_sbm_full}) is important
to deduce the full behaviour of $\overline{\delta^2(\Delta)}$
when $\Delta$ approaches $t$, as shown in Fig.~\ref{fig_msd_cusp} in the Appendix.

\begin{figure}
\includegraphics[width=8cm]{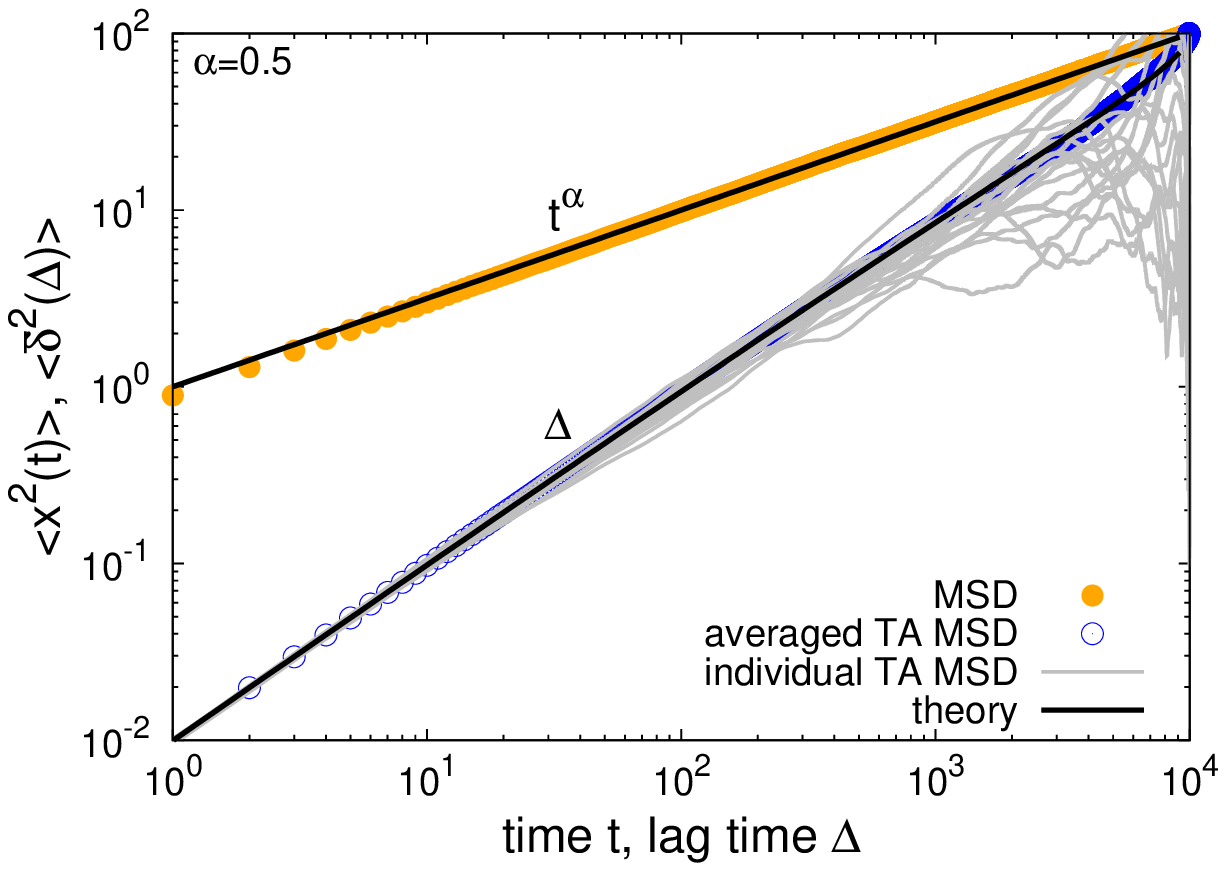}\\[0.2cm]
\includegraphics[width=8cm]{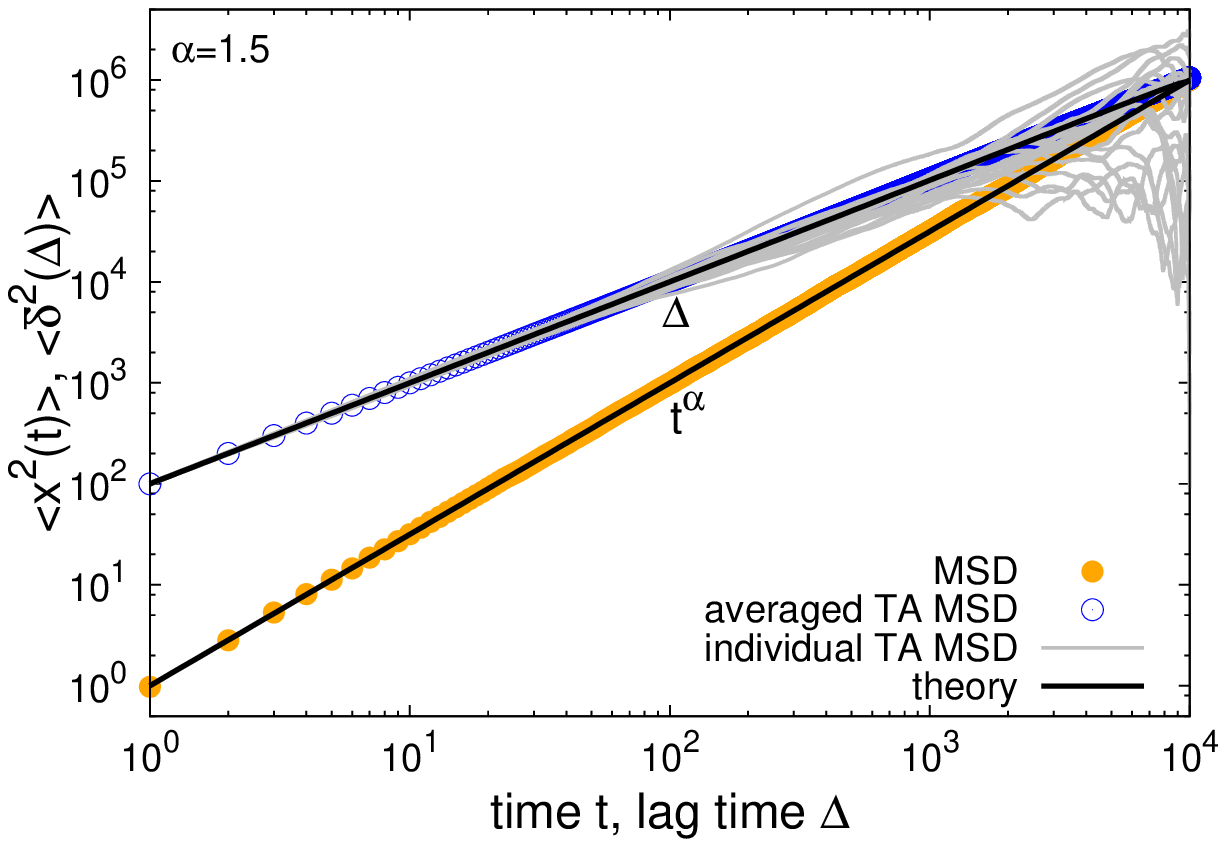}
\caption{MSD and time averaged MSD for SBM with $\alpha=1/2$ (top) and $\alpha=
3/2$ (bottom). The simulations of the SBM-Langevin equation (\ref{langevin})
show excellent agreement with the analytical results, Eqs.~(\ref{msd}) and
(\ref{tamsd_sbm_full}). We also show results for
the time averaged MSD for 20 individual trajectories. Apart from the region where
the lag time $\Delta$ approaches the length $t$ of the time series and statistics
worsen, there is hardly any amplitude scatter between individual $\overline{\delta
^2}$.}
\label{fig_tamsd}
\end{figure}

In Fig.~\ref{fig_tamsd} we show results from simulations of the SBM process for
both sub- and superdiffusion, observing excellent agreement with the analytical
findings. We also see that the scatter between individual trajectories is very
small. Such scatter is a characteristic for anomalous diffusion models and can
be used, e.g., to reliably distinguish FBM from subdiffusive CTRW processes
\cite{pccp,jae_jpa}. For CTRW subdiffusion we find pronounced scatter between
$\overline{\delta^2(\Delta)}$ from individual trajectories even in the limit of
extremely long trajectories $t\to\infty$ \cite{yonghe,pccp,pt}, while for FBM
the scatter vanishes for longer $t$, a characteristic of the ergodic nature of
FBM \cite{goychuk,deng,jae_jpa,pccp}. The amplitude scatter of individual
$\overline{\delta^2}$ can be characterised in terms of the dimensionless variable
$\xi=\overline{\delta^2(\Delta)}/\left<\overline{\delta^2(\Delta)}\right>$. If
$\xi$ has a narrow distribution around $\xi=1$ and the width decreases with
increasing $t$, the process is usually considered ergodic.
This width is characterised in terms of the ergodicity breaking parameter
$\mathrm{EB}=\langle\xi^2\rangle-\langle \xi\rangle^2$. For SBM it was found
\cite{thiel}
\begin{equation}
\label{sbm_eb}
\mathrm{EB}=\left\{\begin{array}{ll}4I_{\alpha}\left(\Delta/t\right)^{2\alpha},
&0<\alpha<1/2\\[0.2cm]
\frac{1}{3}(\Delta/t)\ln(t/\Delta),&\alpha=1/2\\[0.2cm]
\frac{4\alpha^2}{3(2\alpha-1)}\left(\Delta/t\right),&\alpha>1/2
\end{array}\right.
\end{equation}
with the integral $I_{\alpha}=\int_0^1dy\int_0^{\infty}dx[(x+1)^{\alpha}-
(x+y)^{\alpha}]^2$ \cite{thiel}. The $\mathrm{EB}$ parameter for SBM thus
clearly decays to zero for increasing $t$. For $\alpha=1$ we obtain the
known form $\mathrm{EB}=\frac{4}{3}\Delta/t$ for Brownian motion. The
$\Delta/t$ scaling also characterises FBM for $\alpha<3/2$ \cite{deng}. Our
analysis for SBM shows that the scatter distribution $\phi(\xi)$ is indeed narrow
and of Gaussian shape, albeit it is broader than the Gaussian form predicted for
FBM in Ref.~\cite{jae_jpa} (not shown). It decreases with the ratio $\Delta/t$ and
thus indicates a reproducible behaviour between individual trajectories.
The coexistence of the disparity $\overline{\delta^2(\Delta)}
\neq\langle x^2(\Delta)\rangle$ and asymptotically vanishing ergodicity breaking
parameter, $\lim_{t\to\infty}\mathrm{EB}=0$ is a new class of non-ergodic processes
the more detailed mathematical nature of which remains to be examined.

From the analysis so far for free motion we can see that the SBM process has a
truly split personality. Thus its PDF is identical to that for free FBM.
In contrast to the ergodic behaviour $\overline{\delta^2(\Delta)}=\langle x^2(
\Delta)\rangle$ of FBM for sufficiently long $t$ \cite{deng,pccp}, however, SBM
exhibits weak ergodicity breaking, as demonstrated in
Eq.~(\ref{tamsd_sbm}). This scaling form $\overline{\delta^2}\simeq\Delta/t^{1-
\alpha}$ exactly matches the result for
CTRW subdiffusion \cite{pt,yonghe,pccp} or diffusion processes with space-dependent
diffusivity \cite{andrey,andrey_pccp}. Unlike the weakly non-ergodic dynamics of
the latter two, for SBM the fluctuations around the mean $\left<\overline{\delta^2
(\Delta)}\right>$ measured by the distribution $\phi(\xi)$ are narrow and decrease
with longer $t$. We now show that the behaviour of confined SBM is also
unconventional.

\section{Confined SBM}

\begin{figure}
\includegraphics[width=8cm]{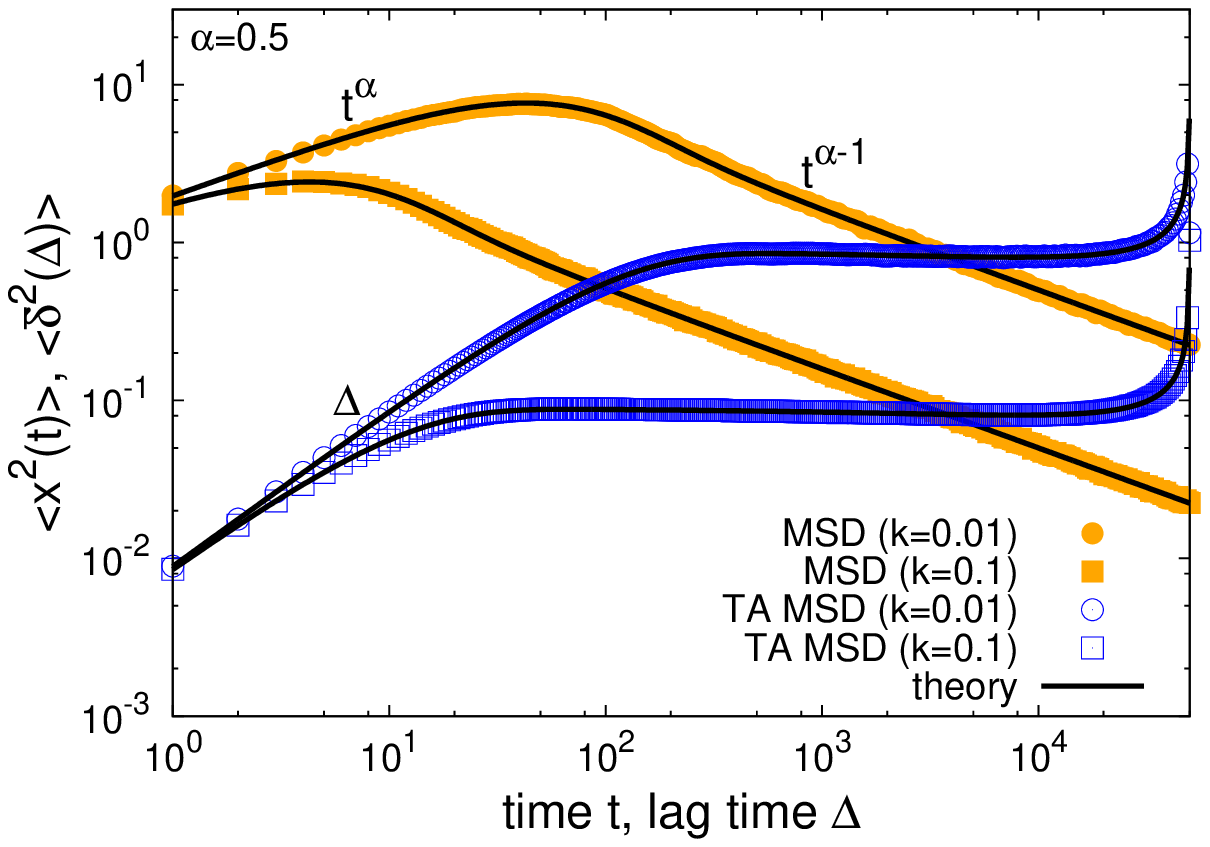}\\[0.2cm]
\includegraphics[width=8cm]{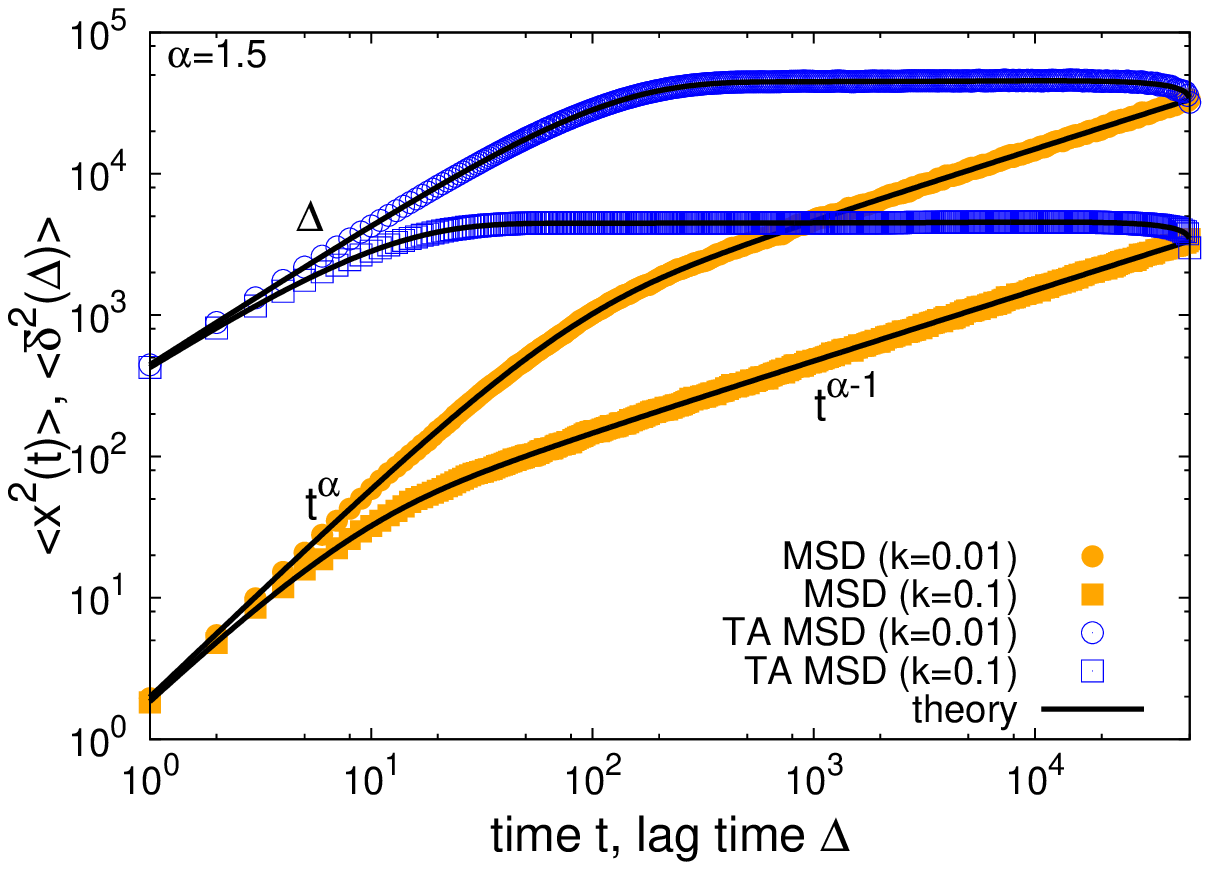}
\caption{MSD $\langle x^2(t)\rangle$ and time averaged MSD $\left<\overline{
\delta^2(\Delta)}\right>$ for $\alpha=1/2$ (top) and $\alpha=3/2$ (bottom) in an
harmonic potential. In each case we consider the force constants $k=0.01$ and
$k=0.1$. Convergence of the corresponding
ensemble and time averages at $t=5\times10^4$ can be shown numerically.
The shown analytical curve is based on the full solution for $\left<\overline{
\delta^2(\Delta)}\right>$ provided in the Appendix.}
\label{harmonic}
\end{figure}

An important physical property of a stochastic process is
its response to external forces or spatial confinement. From an experimental point
of view, this is of relevance to tracer particles moving in the confines of
cellular compartments or when the particle is traced by the help of optical
tweezers, which exert an Hookean restoring force on the particle \cite{lene}.
We study the paradigmatic case of an harmonic potential $V(x)\propto\frac{1}{2}
kx^2$, for which the motion is governed by the Fokker-Planck equation
\begin{equation}
\label{fokker}
\frac{\partial}{\partial t}P(x,t)=\frac{\partial}{\partial x}\left(kx+\mathscr{K}
(t)\frac{\partial}{\partial x}\right)P(x,t),
\end{equation}
which follows directly from the Langevin equation (\ref{langevin}) with the
additional Hookean force term $-kx(t)$. Note that in absence of the external
force ($k=0$) this equation is that of Batchelor \cite{batchelor}. For the MSD
we find an exact expression in terms of the Kummer function $M$ \cite{abramowitz},
\begin{equation}
\langle x^2(t)\rangle=2K_{\alpha}t^{\alpha}e^{-2kt}M(\alpha,1+\alpha,2kt),
\end{equation}
from which we obtain the initial free subdiffusion (\ref{msd}) for $t\ll1/k$ and
the scaling form
\begin{equation}
\label{msd_harm}
\langle x^2(t)\rangle\sim\alpha K_{\alpha}k^{-1}t^{\alpha-1}
\end{equation}
in the long time limit $t\gg1/k$. For subdiffusion ($0<\alpha<1$) the MSD thus
has a power-law decay to zero, while for superdiffusion ($1<\alpha<2$) it
grows infinitely. This
counterintuitive behaviour of SBM is due to the fact that the time dependence of
the diffusivity $\mathscr{K}(t)$ corresponds to a time dependent temperature
\cite{fulinsky} or a time dependent viscosity. Thus SBM is a most non-stationary
process that never reaches stationarity. Fig.~\ref{harmonic} corroborates this
analytical result with simulations based directly on the Langevin equation
(\ref{langevin}): after the free anomalous diffusion behaviour of the MSD for a
particle starting at the vertex of the potential, we observe a turnover to a
power-law behaviour with negative or positive scaling exponent.

The result for the time averaged MSD is similarly remarkable. As shown in
Fig.~\ref{harmonic} for SBM simulations based on the Langevin equation
(\ref{langevin}) with the Hookean forcing, it exhibits a pronounced apparent
plateau for lag times $\Delta\gg1/k$ for both sub- and superdiffusive SBM.
This behaviour is in excellent agreement with the full analytical solution
(\ref{tamsd_harm_full}) provided in the Appendix in terms of Kummer functions.
Taking the limit $\Delta\ll t$ we obtain
\begin{equation}
\label{tamsd_harm}
\left<\overline{\delta^2(\Delta)}\right>\sim\frac{K_{\alpha}}{k}\left[\frac{t^{
\alpha}-\Delta^{\alpha}}{t-\Delta}+(t-\Delta)^{\alpha-1}\left(1-2e^{-k\Delta}
\right)\right],
\end{equation}
which indeed features the extended plateau and provides a good approximation
for $\Delta\gg1/k$. Note that when $\Delta$ approaches the measurement time $t$
the time averaged MSD $\left<\overline{\delta^2(\Delta)}\right>$ converges to the
value of the MSD $\langle x^2(t)\rangle$ due to the pole in expression
(\ref{tamsd}). This behaviour is analysed in more detail in
Fig.~\ref{fig_msd_cusp} in the Appendix.

\begin{figure}
\hspace*{-0.8cm}\includegraphics[width=9.8cm]{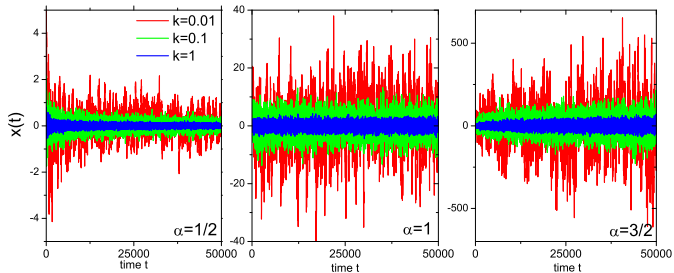}
\caption{Trajectories of SBM in an harmonic potential, for $\alpha=1/2$ (left),
$\alpha=1$ (centre), and $\alpha=3/2$ (right) for three different confinement
strengths $k$. The fluctuations are stationary only in the Brownian case
$\alpha=1$.}
\label{harmtraj}
\end{figure}

Fig.~\ref{harmtraj} analyses this behaviour in the harmonic potential further by
showing the time series $x(t)$ for subdiffusive SBM, Brownian motion, and
superdiffusive SBM. Indeed we see that for subdiffusive and superdiffusive confined
SBM the fluctuations continue to decrease and increase with time, while those in
the Brownian limit become stationary. This is the direct effect of the time
dependent temperature or viscosity encoded in the SBM diffusivity $\mathscr{K}(t)$.
If the system is stationary or thermalised this behaviour clearly
underlines the unsuitability of SBM for the description of anomalous diffusion.

The behaviour of SBM dynamics under confinement is the central result of our
study. The continued temporal decay or increase of the MSD that we obtained for
SBM is in stark contrast to the behaviour of confined CTRW subdiffusion, for which
the MSD
$\langle x^2\rangle$ saturates to the thermal plateau, while the time averaged
MSD continues to grow in the power-law form $\overline{\delta^2}\simeq(\Delta/t)
^{1-\alpha}$ up until the lag time $\Delta$ approaches $t$ \cite{pnas,pccp}. It
strongly differs from FBM, which relaxes to a plateau for both the MSD and the
time averaged MSD, and for which only a transient disparity between $\langle x^2
\rangle$ and $\overline{\delta^2}$ exists \cite{lene1}. Finally, SBM is also at
variance with heterogeneous
diffusion processes with a space dependent diffusion coefficient relax to a plateau
for both $\langle x^2\rangle$ and $\overline{\delta^2}$ \cite{andrey_confined}.

\section{Discussion}

As we showed SBM is a truly paradoxical stochastic process. Somewhat similar to
a chameleon, each time we compare SBM with other established anomalous diffusion
processes we find certain similarities. Looking at the sum of its features,
however, SBM is a truly independent process with a range of remarkable properties.

For free SBM the probability density function $P(x,t)$ equals that of FBM, despite
the fact that both processes are governed by different stochastic (Langevin)
equations. SBM's time averaged MSD scales equally to those of the weakly
non-ergodic CTRW subdiffusion and diffusion processes with space-dependent
diffusivity. Despite this weakly non-ergodic character of the mean time
averaged MSD $\left<\overline{\delta^2}\right>$, the amplitude scatter between
the time averaged MSD $\overline{\delta^2}$ of individual realisations is
small and the distribution has a Gaussian shape, as otherwise observed for the
ergodic FBM or for Brownian motion of finite time $t$. In that sense SBM
represents a new class of non-ergodic processes. The most striking behaviour of
SBM is, however, its strongly non-stationary behaviour under confinement. Instead
of relaxing to a plateau the MSD acquires a power-law decay or growth
mirroring a continuously decreasing or increasing temperature encoded in
SBM's time dependent diffusion coefficient $\mathscr{K}(t)$.

SBM is thus at variance with the currently available experimental observations in
complex liquids using single particle tracking by video microscopy or by optical
tweezers tracking of single submicron particles. Thus the free anomalous diffusion
data garnered so far was classified into FBM-like and CTRW-like behaviour, or
combinations thereof \cite{weber,weigel,tabei}. Note that also for fluorescence
correlation spectroscopy experiments recent analysis tools corroborated  an FBM
nature of the data \cite{szymanski}. For optical tweezers tracking of lipid
granules in different complex liquids the time averaged MSD either
continues to grow under confinement, reflecting the non-ergodic features of CTRWs
\cite{lene,pt}, or it relaxes to a plateau value mirroring an ergodic dynamics
\cite{lene1,naturephot}.

How can FBM and SBM have the same distribution (\ref{prop}) in free space? Simply
put for FBM, the viscoelastic properties of the environment effecting the
long-range correlations of FBM lead to a frequency dependent response of the
environment to a disturbance, while the materials properties remain unchanged in
time. For free FBM this gives rise to the subdiffusive MSD
(\ref{msd}), in which the time dependent diffusivity effectively encodes the
frequency dependent response of the viscoelastic environment. The distribution
(\ref{prop}) for FBM and its description in terms of a Fokker-Planck equation of
the form (\ref{fokker}) is treacherous, however. This can already be seen when
we use the PDF (\ref{prop}) or the dynamics equation (\ref{fokker}) to
calculate the first passage behaviour. This procedure leads to the wrong result
\cite{epl}, and the full analytical description of FBM in the presence of
boundaries remains elusive, a difficulty imposed by the highly correlated
fractional Gaussian noise driving its Langevin equation. In
contrast, SBM, according to its Langevin description (\ref{langevin}),
is driven by uncorrelated noise but the environment itself is changing as
function of time, effecting an extremely non-stationary process. The equivalence
of the PDF (\ref{prop}) of both processes is thus simply due to the fact that a
Gaussian PDF is completely defined by its second moment, the MSD (\ref{msd}).

With its interesting behaviour that is so different from the other conventional
anomalous diffusion models, SBM may indeed
have relevant applications to weakly coupled or fully adiabatic systems, as
well as for active systems, in which the existence of a temperature is not
meaningful. In particular, in the superdiffusive case SBM or analogous dynamics
with other increasing effective diffusivities may represent an alternative
approach to active Brownian motion \cite{ebeling}.

\begin{figure}
\includegraphics[width=8cm]{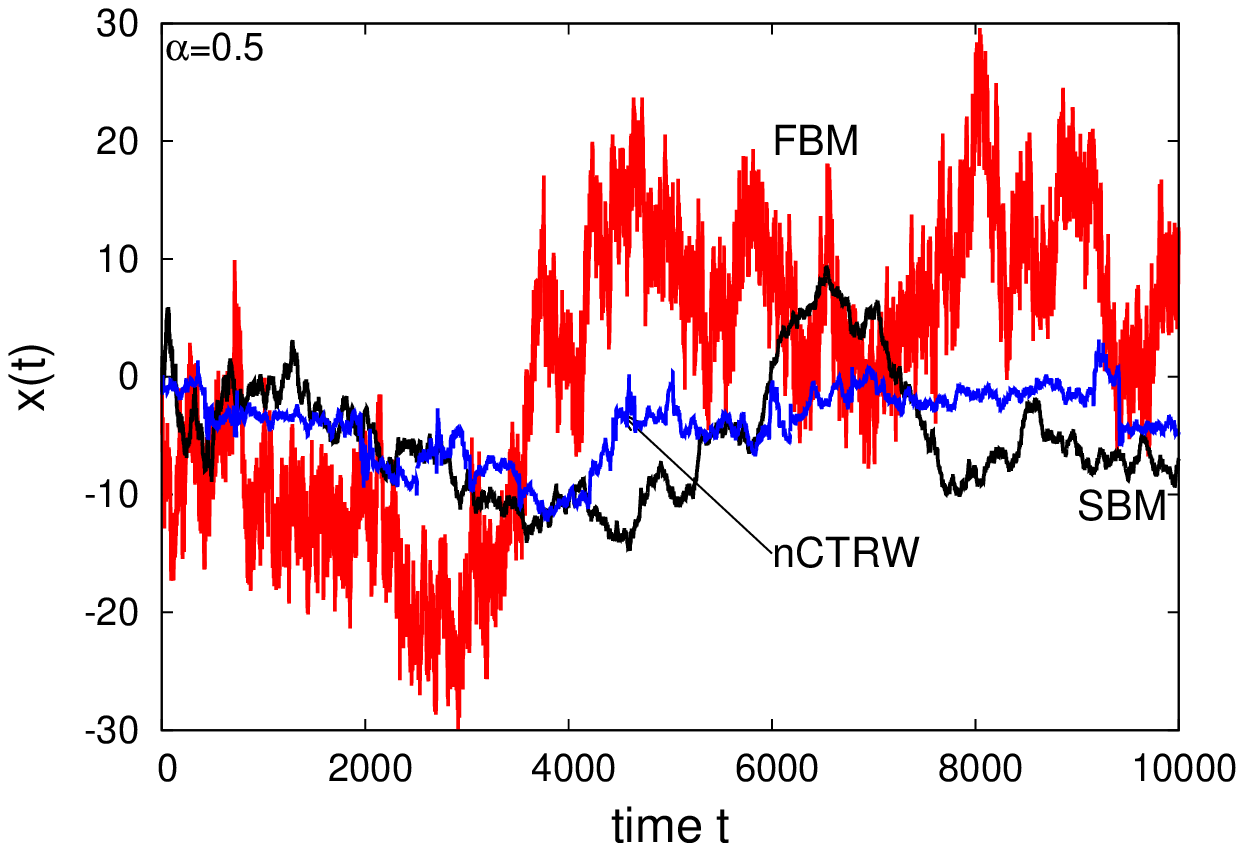}\\[0.2cm]
\includegraphics[width=8cm]{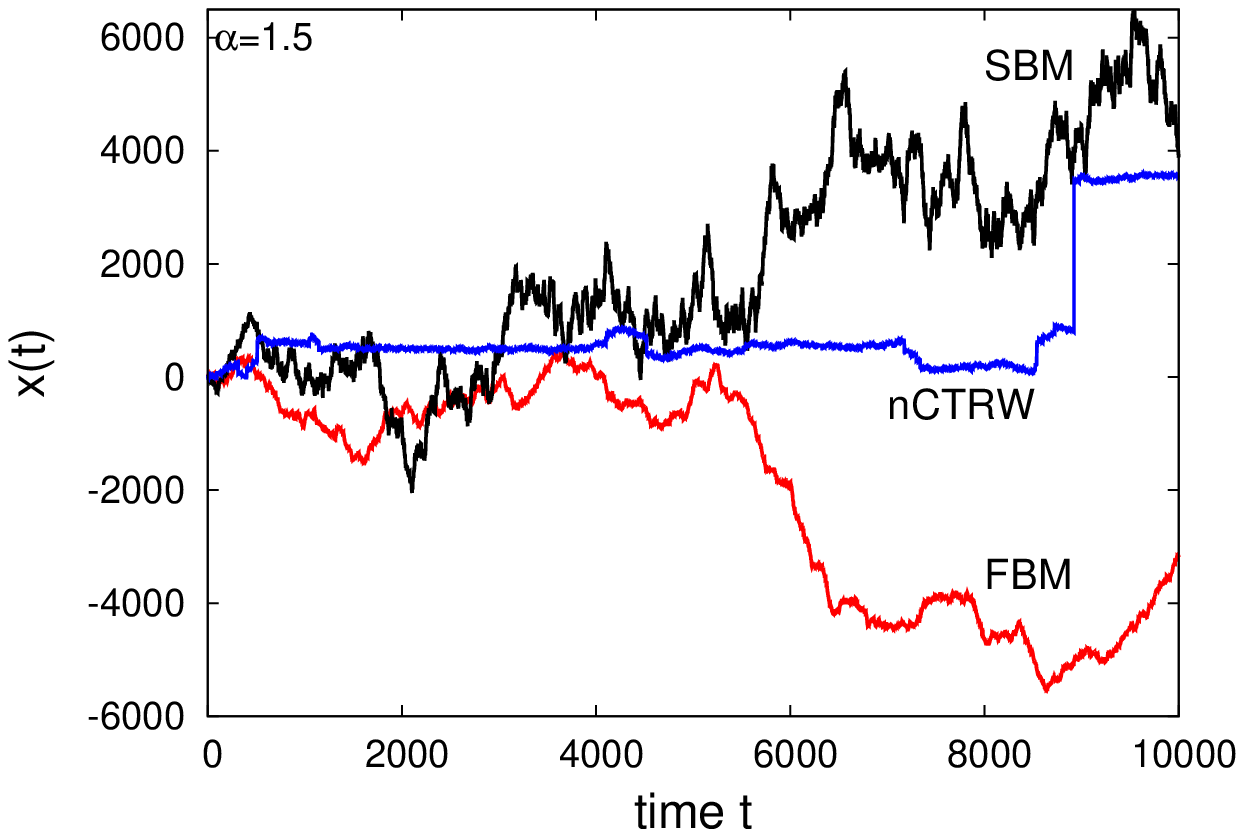}
\caption{Individual trajectories of anomalous stochastic processes: FBM, SBM,
and noisy CTRW (nCTRW), for subdiffusion with $\alpha=1/2$ (top) and
superdiffusion with $\alpha=3/2$ (right).}
\label{trajs}
\end{figure}

The difference of SBM to other processes can also be seen in Fig.~\ref{trajs}.
For the subdiffusive case a sample trajectory of SBM is compared to that of FBM and
the noisy CTRW \cite{noisy}, in which the pure subdiffusive CTRW is superimposed
with Ornstein-Uhlenbeck noise to accommodate the thermal noise of the environment
observed in experiments \cite{wong}. We see that SBM with its Gaussian probability
density function and uncorrelated driving noise appears more similar to the noisy
CTRW motion, albeit
it has a more pronounced tendency to reach larger amplitudes than the CTRW with
its waiting time periods. The fluctuations of SBM are dramatically less than
those of the anticorrelated FBM, which also shows the largest amplitudes in
$x(t)$. For superdiffusion, we compare SBM with a noisy version of the L{\'e}vy
walk process \cite{jeocheme} and again with FBM. Note the vastly different size
of the window shown on the ordinate. This time, the persistent FBM shows
pronouncedly larger excursions and a distinct reduction of the noise compared to
SBM. The shape of $x(t)$ of the latter does not appear to be qualitatively
different from the
subdiffusive case. The noisy CTRW is fundamentally different from both SBM and FBM.
Despite their disparate physical nature and their dissimilarity when setting
one against the other in a direct comparison, we note that generally it is
difficult to identify a stochastic process solely from the appearance of the
recorded trajectory.

From this discussion it is obvious that one thing is to have at our disposal a
handy and easy-to-use description for anomalous diffusion processes. SBM with
its Gaussian and uncorrelated nature appears deceivingly simple and is therefore
easy to implement in numerical analyses and descriptions such as
diffusion limited reactions. However, the other face of the medal is about the
physical relevance of a model process. Using SBM with its time-dependent diffusion
coefficient violates the physical setting in typical experiments, in which the
system is held at approximately constant temperature and its predictions are at
odds with actual observations. The unphysical nature for the kind of processes we
have in mind is most obvious for confined motion. In contrast, FBM and fractional 
Langevin equation motion are ergodic processes with the physical background of an
effective particle motion in a viscoelastic multi-body environment. FBM and
fractional Langevin equation motion exhibit a transient disparity between the MSD
and the time averaged MSD under confinement \cite{lene1}. Weakly non-ergodic
CTRW processes emerge due to immobilisation periods imprinted on the
dynamics by the structure of the environment of binding events to the environment.
Finally, diffusion processes with space dependent diffusivities
arise naturally in non-homogeneous systems such as biological cells or subsurface
aquifers.
To extract physically meaningful information from anomalous diffusion
data one needs to have some physical insight into the observed process and properly
analyse the data using complementary tools \cite{pt,tabei,weigel,garini,lene1,
noisy,vincent,radons,berez} before settling for the appropriate physical model.

We finally note that it will be interesting to compare the predictions of the SBM
model with that of active processes in viscoelastic environments \cite{active,
active1}.

\acknowledgments

We acknowledge funding from the Academy of Finland within the FiDiPro scheme.\\

\begin{appendix}

\section*{Appendix}

The covariance of the position for SBM in an harmonic potential follows from the
Langevin equation (\ref{langevin}). Our result is
\begin{equation}
\label{cova}
\left< x(t_1)x(t_2)\right>=2K_{\alpha}t_1^{\alpha}e^{-k(t_1+t_2)}M(\alpha,1+
\alpha,2kt_1)
\end{equation}
for $t_1<t_2$. In absence of the confinement ($k=0$) the covariance reduces to
the MSD (\ref{msd}) and in the Brownian limit $\alpha=1$ we recover the familiar
covariance
\begin{equation}
\left< x(t_1)x(t_2)\right>=\frac{K_1}{k}\left(e^{-k(t_2-t_1)}-e^{-k(t_1+t_2)}
\right).
\end{equation}
From Eq.~(\ref{cova}) we derive the exact result for the time averaged MSD
(\ref{tamsd}),
\begin{eqnarray}
\nonumber
\left<\overline{\delta^2(\Delta)}\right>&=&\frac{2K_{\alpha}}{1+\alpha}
(t-\Delta)^{\alpha}e^{-2k(t-\Delta)}\\
\nonumber
&&\times M(1+\alpha,2+\alpha,2k[t-\Delta])\\
\nonumber
&&\hspace*{-1.2cm}
+\frac{2K_{\alpha}}{(1+\alpha)(t-\Delta)}\Big[t^{1+\alpha}e^{-2kt}M(1+\alpha,
2+\alpha,2kt)\\
\nonumber
&&\hspace*{0.6cm}-\Delta^{1+\alpha}e^{-2k\Delta}M(1+\alpha,2+\alpha,2k\Delta)\Big]\\
\nonumber
&&\hspace*{-1.2cm}
-\frac{4K_{\alpha}}{1+\alpha}(t-\Delta)^{\alpha}e^{-2kt+k\Delta}\\
&&\times M(1+\alpha,2+\alpha,2k[t-\Delta]).
\label{tamsd_harm_full}
\end{eqnarray}
To derive the limit (\ref{tamsd_harm}) of the time averaged MSD for confined SBM
in the long time limit $t\to\infty$ we use the property
\begin{equation}
M(\alpha,1+\alpha,x)\sim\frac{\Gamma(1+\alpha)}{\Gamma(\alpha)}\times\frac{e^x}{x}
\end{equation}
of the Kummer function \cite{abramowitz}.

\begin{figure}
\includegraphics[width=8cm]{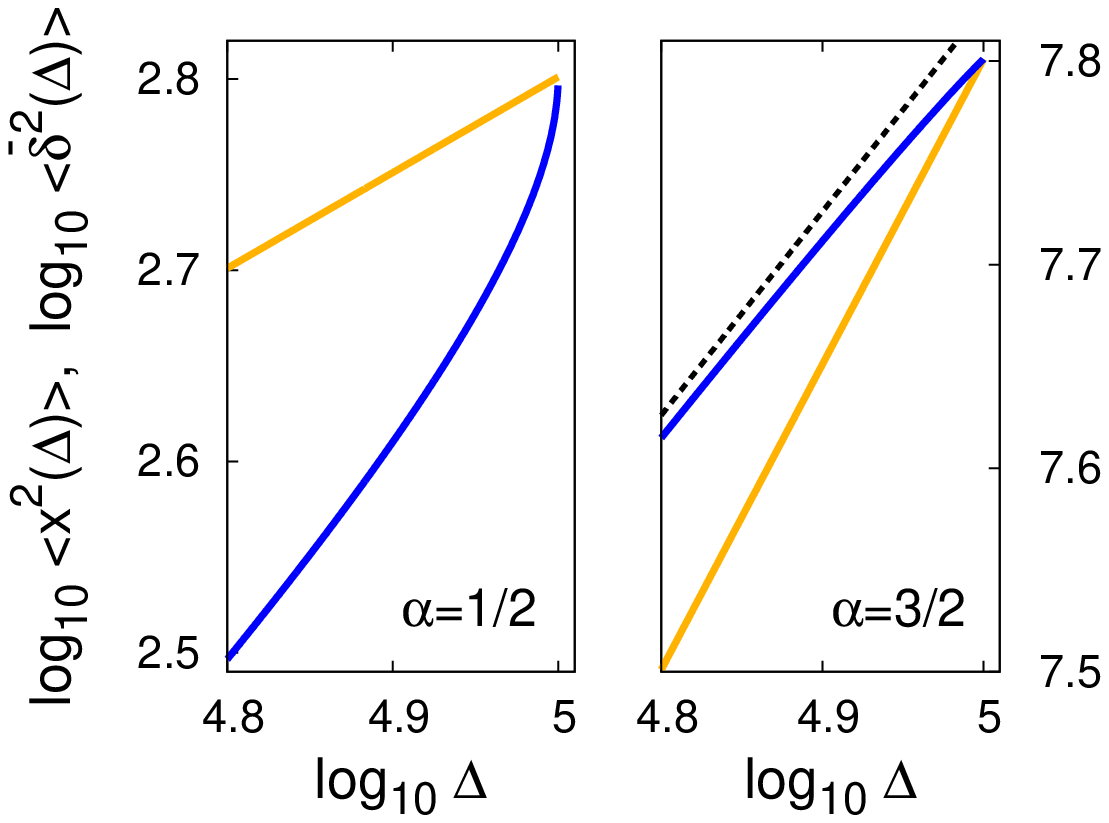}\\[0.2cm]
\includegraphics[width=8cm]{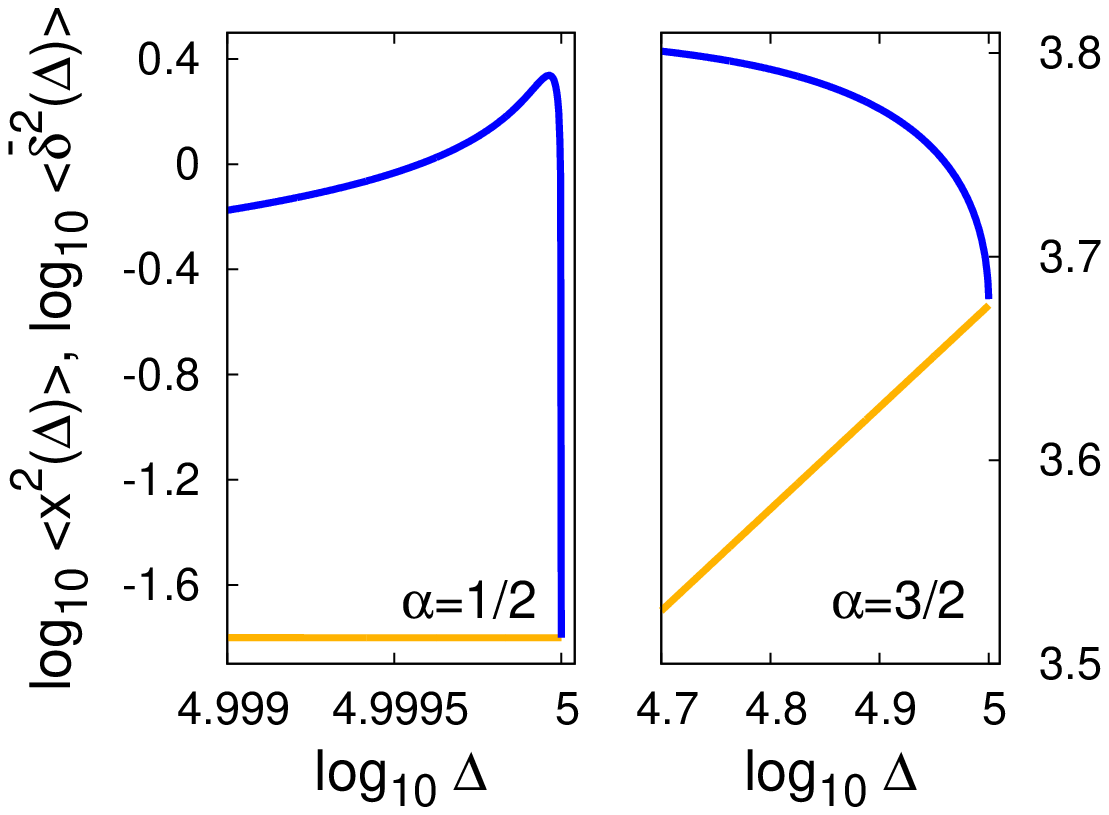}
\caption{Convergence of the time averaged MSD $\left<\overline{\delta^2(\Delta)}
\right>$ to the MSD $\langle x^2(\Delta)\rangle$ at $\Delta\to t$ for $\alpha=1/2$
and $\alpha=3/2$. The measurement time $t=10^5$. Top: Free SBM. For the
superdiffusive case the dashed black line of unity slope aids in demonstrating
the non-linear behaviour of $\left<\overline{\delta^2(\Delta)}\right>$. Bottom: SBM
in an harmonic potential with $k=0.1$. Note the different ranges of the abscissa
in the two cases. For the extremely small window for the subdiffusive case needed
to illustrate the cusp at $\Delta\to t$ and the relatively large variation of
$\left<\overline{\delta^2(\Delta)}\right>$ the MSD $\langle x^2(\Delta)\rangle$
appears almost constant.}
\label{fig_msd_cusp}
\end{figure}

In Fig.~\ref{fig_msd_cusp} we illustrate the convergence of the time averaged
MSD $\left<\overline{\delta^2(\Delta)}\right>$ to the value of the MSD $\langle
x^2(t)\rangle$ in the limit $\Delta\to t$.

\end{appendix}

\end{document}